\newcommand{\openone}{\leavevmode\hbox{\small1\normalsize\kern-.33em1}}
\newcommand{\R}{\mathop{\mathrm{Re}}\nolimits}
\newcommand{\I}{\mathop{\mathrm{Im}}\nolimits}

\documentclass[10pt]{iopart}
\usepackage{iopams,bm,cite,graphicx}

\begin{document}

\title{Geometrical aspects
of first-order optical systems}

\author{A. G. Barriuso{}$^{\mathrm{1}}$,
J. J. Monz\'on{}$^{\mathrm{1}}$,
L. L. S\'{a}nchez-Soto{}$^{\mathrm{1}}$ and
J.~F.~Cari\~{n}ena{}$^{\mathrm{2}}$}

\address{{}$^{\mathrm{1}}$ Departamento de \'Optica,
Facultad de F\'{\i}sica, Universidad Complutense,
28040~Madrid, Spain}

\address{{}$^{\mathrm{2}}$ Departamento de
F\'{\i}sica Te\'orica, Facultad de Ciencias,
Universidad de Zaragoza, 50009 Zaragoza, Spain}

\eqnobysec

\begin{abstract}
We reconsider the basic properties of ray-transfer
matrices for first-order optical systems from
a geometrical viewpoint. In the paraxial regime
of scalar wave optics, there is a wide family of
beams for which the action of a ray-transfer matrix
can be fully represented as a bilinear transformation
on the upper complex half-plane, which is the
hyperbolic plane. Alternatively, this action
can be also viewed in the unit disc. In both
cases, we use a simple trace criterion that
arranges all first-order systems in three classes
with a clear geometrical meaning: they represent
rotations, translations, or parallel displacements.
We analyze in detail the relevant example of
an optical resonator.

\bigskip

\end{abstract}


\section{Introduction}

Matrix methods~\cite{HK71,Bar90} offer the
great advantage of simplifying the presentation
of linear models and clarifying the common
features and interconnections of distinct
branches of physics~\cite{Kau94}. Modern
optics is not an exception and a wealth
of input-output relations can be compactly
expressed by a single matrix~\cite{GB75}.
For example, the well-known $2 \times 2$
ray-transfer matrix, which belongs to the
realm of paraxial ray optics, predicts with
almost perfect accuracy the behavior of a
Gaussian beam.

In this respect, we note that there is a wide
family of beams (including Gaussian Schell-model
fields, which have received particular
attention~\cite{WC78,FZ78,Sal79,Gor80,SW82,FS82,Gor83,GG84,FT88,FTT94,ABG94})
for which a complex parameter can be
defined such that, under the action
of first-order systems, it is transformed
according to the famous Kogelnik $ABCD$
law~\cite{Col63,Li64,Kog65a,Kog65b,KL66}. This
is the reason why they are so easy to handle.
This simplicity, together with the practical
importance that these beams have for laser
systems, explain the abundant literature on
this topic~\cite{Sie86,ST91}.

The algebraic basis for understanding
the transformation properties of
such beams is twofold: the ray-transfer
matrix of any first-order system is
an element of the group
SL(2, $\mathbb{R}$)~\cite{Gil74} and
the complex beam parameter changes
according to a bilinear (or M\"{o}bius)
transformation~\cite{Kr90}.

The nature of these results seems to call
for a geometrical interpretation. The
interaction between physics and geometry
has a long and fruitful story, a unique
example is Einstein theory of relativity.
The goal of this paper is precisely to provide
such a geometrical basis, which should be relevant
to properly approach this subject.

The material of this paper is organized
as follows. In section~\ref{halfplane} we include
a brief review of the transformation
properties of Gaussian beams by first-order
systems, introducing a complex parameter $Q$
to describe the different states as points of
the hyperbolic plane. The action of the system
in terms of $Q$ is then given by a bilinear
transformation, which is characterized through
the points that it leaves invariant. From
this viewpoint the three basic isometries of
this hyperbolic plane (i.e., transformations
that preserve the distance), namely, rotations,
translations, and parallel displacements, appear
linked to the fact that the trace of the ray-transfer
matrix has a magnitude lesser than, greater than,
or equal to 2, respectively.

In section~\ref{unitdisc} we present a mapping
that transforms the hyperbolic plane into the
unit disc (which is the Poincar\'e model of the
hyperbolic geometry) and we proceed to study the
corresponding motions in this disc. Finally, as
a direct application, in section~\ref{Periodic}
we treat the case of periodic systems, which are
the basis for optical resonators, providing an
alternative explanation of the standard stability
condition.

We emphasize that this geometrical scenario
does not offer any advantage in terms of
computational efficiency. Apart from its
undeniable beauty, its benefit lies in
gaining insights into the qualitative
behaviour of the beam evolution.

\section{First-order systems as transformations
in the hyperbolic plane $\mathbb{H}$}

\label{halfplane}

We consider the paraxial propagation of
light through axially symmetric systems,
containing no tilted or misaligned elements.
The reader interested in further details
should consult the extensive work of Simon
and Mukunda~\cite{SSM84,SSM85,SMS88,SM93,SM98}.
We take a Cartesian coordinate system whose
$Z$ axis is along the axis of the optical
system and represent a ray at a plane $z$
by the transverse position vector $x (z)$
(which can be chosen in the meridional plane)
and by the momentum $ p (z) = n (z)
dx / dz$~\cite{KBW04}. Here $n(z)$ is
the refractive index and $d x/dz$ is
the direction of the ray through $z$.

At the level of ray optics, a first-order
system changes the ray parameters by the
simple transformation~\cite{BGK04}
\begin{equation}
\label{Mgeo}
\left (
\begin{array}{c}
x^\prime \\
p^\prime
\end{array}
\right ) =
\mathbf{M}
\left (
\begin{array}{c}
x \\
p
\end{array}
\right ) ,
\end{equation}
where the primed and unprimed variables refer
to the output and input planes, respectively,
and $\mathbf{M}$ is the ray-transfer matrix
that must satisfy the condition~\cite{SW00}
\begin{equation}
\mathbf{M} =
\left (
\begin{array}{cc}
A & B \\
C & D
\end{array}
\right ) ,
\qquad
\det \mathbf{M} = AD - BC = 1 ,
\end{equation}
which  means that $\mathbf{M}$ is an element
of the group SL(2, $\mathbb{R}$) of real
unimodular $2 \times 2$ matrices.

When one goes to paraxial-wave optics, the
beams are described in the Hilbert space
$L^2$ of complex-valued square-integrable
wave-amplitude functions $\psi (x)$. The
classical phase-space variables $x$ and $p$
are now promoted to self-adjoint operators
by the procedure of wavization~\cite{GM69},
which is quite similar to the quantization
of position and momentum in quantum mechanics.

We are interested in the action of a
ray-transfer matrix on time-stationary
fields. We can then focus the analysis
on a fixed frequency $\omega$, which
we shall omit henceforth. Moreover,
to deal with partially coherent beams
we specify the field not by its amplitude,
but by its cross-spectral density. The latter
is defined in terms of the former as
\begin{equation}
\label{Gamma}
\Gamma (x_1, x_2) =
\langle \psi^\ast (x_1)
\psi (x_2) \rangle ,
\end{equation}
where the angular brackets denote ensemble
averages.

There is a wide family of beams, known as
Schell-model fields, for which the
cross-spectral density (\ref{Gamma})
factors in the form
\begin{equation}
\Gamma (x_1, x_2) =
[I(x_1)I(x_2)]^{1/2}
\mu (x_1 - x_2) ,
\end{equation}
where $I$ is the intensity distribution and
$\mu$ is the normalized degree of coherence,
which is translationally invariant. When
these two fundamental quantities are Gaussians
\begin{eqnarray}
I(x) & = & \frac{\mathcal{I}}
{\sqrt{2 \pi} \sigma_I}
\exp \left ( - \frac{x^2}{2 \sigma_I^2} \right ) ,
\nonumber \\
& & \\
\mu (x) & = &
\exp \left ( - \frac{x^2}
{2 \sigma_\mu^2} \right ) ,
\nonumber
\end{eqnarray}
the beam is said to be a Gaussian Schell model
(GSM). Here $\mathcal{I}$ is a constant independent
of $x$ that can be identified with the total
irradiance. Clearly, $\sigma_I$ and $\sigma_\mu$
are, respectively, the effective beam width and
the transverse coherence length. Other well-known
families of Gaussian fields are special cases of
these GSM fields. When $\sigma_\mu \ll \sigma_I$
we have the Gaussian quasihomogeneous field,
and the coherent Gaussian field is obtained
when $\sigma_\mu \rightarrow \infty$. In any
case, the crucial point for our purposes is
the observation that for GSM fields one can
define a complex parameter $Q$~\cite{Dra96b}
\begin{equation}
\label{Qpar}
Q =  \frac{1}{R} + i \frac{1}
{k \, \sigma_I \; \delta} ,
\end{equation}
where
\begin{equation}
\frac{1}{\delta^2} =
\frac{1}{\sigma_\mu^2}
+ \frac{1}{(2 \sigma_I)^2} ,
\end{equation}
and $R$ is the wave front curvature radius.
This parameter fully characterizes the beam
and satisfies the Kogelnik $ABCD$ law;
namely, after propagation through a
first-order system, the parameter $Q$
changes to $Q^\prime$ via
\begin{equation}
\label{MOBQ}
Q^\prime =  \Psi [\mathbf{M}, Q ] =
\frac{C + D  Q}
{A + B Q} .
\end{equation}
Since $\I Q > 0$ by the definition
(\ref{Qpar}),  one immediately checks that
$\I Q^\prime > 0$ and we can thus
view the action of the first-order system
as a bilinear transformation $\Psi$ on
the upper complex half-plane. When we use the
metric $ds = |dQ|/\I Q$ to measure distances,
what we get is the standard model of the
hyperbolic plane $\mathbb{H}$~\cite{Sta93}.
This plane $\mathbb{H}$ is invariant under bilinear
transformations.

We note that the whole real axis, which is
the boundary of $\mathbb{H}$, is also invariant
under (\ref{MOBQ}) and represents wave fields
with unlimited transverse irradiance (contrary
to the notion of a beam). On the other hand,
for the points in the imaginary axis we have
an infinite wave front radius, which defines
the corresponding beam waists. The origin
represents a plane wave.

Bilinear transformations constitute an
important tool in many branches of physics.
For example, in polarization optics they
have been employed for a simple
classification of polarizing devices
by means of the concept of eigenpolarizations
of the transfer function~\cite{AB87,HKN96}.

In our context, the equivalent concept can be
stated as the beam configurations such
that $Q = Q^\prime$ in equation~(\ref{MOBQ}),
whose solutions are
\begin{equation}
Q_{\pm} = \frac{1}{2B} \left [ (D-A) \pm
\sqrt{(A + D)^2-4} \right ] .
\end
{equation}
These values of $Q$ are known as the fixed points
of the transformation.

The trace of $\mathbf{M}$, $\Tr (\mathbf{M}) =
A + D$, provides a suitable tool for the
classification of optical systems~\cite{SMY01}.
It has also played an important role in studying
propagation in periodic media~\cite{Lek94}.
When $ [\Tr ( \mathbf{M} )] ^2 < 4$ the action
is said elliptic and there are no real roots:
they are complex conjugates and  only one of
them lies in $\mathbb{H}$, while the other lies
outside.  When $ [ \Tr ( \mathbf{M} )]^2 > 4$
there are two real roots (i.e., in the
boundary of $\mathbb{H}$) and the action is
hyperbolic. Finally, when $ [ \Tr ( \mathbf{M} )]^2
= 4$ there is one (double) real solution
and the system action is called parabolic.

To proceed further let us  note that by taking
the conjugate of $\mathbf{M}$ with any matrix
$\mathbf{C} \in $ SL($2, \mathbb{R})$
\begin{equation}
\label{conjC}
\mathbf{M}_{\mathrm{C}} = \mathbf{C} \,
\mathbf{M} \, \mathbf{C}^{-1} ,
\end{equation}
we obtain another matrix of the same type,
since $\Tr ( \mathbf{M} ) = \Tr (
\mathbf{M}_{\mathrm{C}} )$. Conversely, if
two systems have the same trace, one can
always find a matrix $\mathbf{C}$ satisfying
equation~(\ref{conjC}).

Note that $Q$ is a fixed point of $\mathbf{M}$
if and only if the image of $Q$ by $\mathbf{C}$
(i.e., $\Psi[\mathbf{C}, Q]$) is a fixed point of
$\mathbf{M}_{\mathrm{C}}$. In consequence,
given any ray-transfer matrix $\mathbf{M}$ one
can always find a $\mathbf{C}$ such that
$\mathbf{M}_{\mathrm{C}}$ takes one of the
following canonical forms~\cite{YMS02,MYS02}:
\begin{eqnarray}
\label{Iwasa1}
\mathbf{K}_{\mathrm{C}} (\vartheta ) & = &
\left (
\begin{array}{cc}
\cos ( \vartheta/2) & \sin (\vartheta/2) \\
- \sin (\vartheta/2) & \cos ( \vartheta/2)
\end{array}
\right ) \ ,
\nonumber \\
\mathbf{A}_{\mathrm{C}} (\xi) & = &
\left (
\begin{array}{cc}
e^{\xi/2}  & 0 \\
0  & e^{-\xi/2}
\end{array}
\right ) \ , \\
\mathbf{N}_{\mathrm{C}} ( \nu ) & = &
 \left (
\begin{array}{cc}
1 & 0 \\
\nu & 1
\end{array}
\right ) \ ,
\nonumber
\end{eqnarray}
where $ 0 \le \vartheta \le 4 \pi$ and
$\xi, \nu \in \mathbb{R}$. These matrices
define the one-parameter subgroups of
SL(2, $\mathbb{R}$) and have as
fixed points $+i$ (elliptic), 0 and $ \infty$
(hyperbolic), and $ \infty$ (parabolic),
respectively. They are the three basic blocks
in terms of which any system action can
be expressed. Clearly,  $\mathbf{K}_{\mathrm{C}}
(\vartheta )$ represents a rotation in phase
space, $\mathbf{A}_{\mathrm{C}} (\xi)$ is a
magnifier that scales $x$ up by the
factor $e^{\xi/2}$ and $p$ down by
the same factor, and $\mathbf{N}_{\mathrm{C}}
( \nu )$ represents the action of a thin
lens of power $\nu$ (i.e., focal length
$1/\nu$)~\cite{KBW04}.

For the canonical forms (\ref{Iwasa1}), the
corresponding actions are
\begin{eqnarray}
\label{QactH}
Q^\prime & = &
\frac{\cos (\vartheta/2) Q -
\sin (\vartheta/2)}
{\sin (\vartheta/2) Q  +
\cos (\vartheta/2)} , \nonumber \\
& & \nonumber \\
Q^\prime & = & e^{- \xi} Q , \\
& & \nonumber \\
Q^\prime & = & Q + \nu . \nonumber
\end{eqnarray}
The first is a rotation, in agreement with
Euclidean geometry, since a rotation has
only one invariant point. The second is a
translation because it has no fixed points
in $\mathbb{H}$ and the geodesic line
joining the two fixed points  (0 and
$\infty$) remains invariant (it is the
axis of the translation). The third
one is known as a parallel displacement.

\begin{figure}
\centering
\resizebox{0.75\columnwidth}{!}{\includegraphics{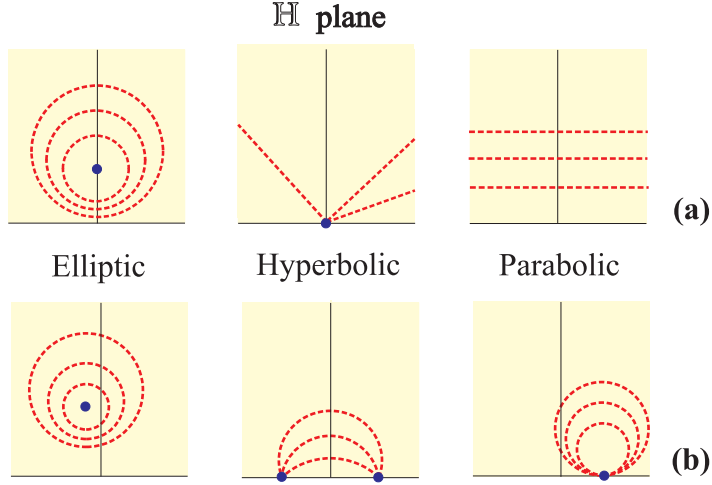}}
\caption{Plot of typical orbits in the hyperbolic
plane $\mathbb{H}$: (a) canonical transfer
matrices as given in equation~(\ref{Iwasa1}) and
(b) arbitrary transfer matrices obtained by
matrix conjugation as in equation~(\ref{conjC}).}
\end{figure}

When one of the parameters $\theta$, $\xi$, or
$\nu$ in (\ref{QactH}) varies, the transformed points
$Q^\prime$ describe a curve called the orbit
of $Q$ under the action of the corresponding
one-parameter subgroup. In figure~1.a
we have plotted typical orbits for
the canonical forms (\ref{Iwasa1}). For matrices
$\mathbf{K}_{\mathrm{C}}(\vartheta)$ the orbits
are circumferences centered at the invariant
point $+ i$ and passing through $Q$ and $-1/Q$.
For $\mathbf{A}_{\mathrm{C}} (\xi)$, they are
lines going from 0 to the $\infty$ through $Q$
and they are known as hypercicles. Finally,
for matrices $\mathbf{N}_{\mathrm{C}} (\nu)$
the orbits are lines parallel to the real axis
passing through $Q$ and they are known as
horocycles~\cite{Cox69}.

For a general matrix $\mathbf{M}$ the
corresponding orbits can be obtained by
transforming with the appropriate matrix
$\mathbf{C}$ the orbits described before.
The explicit construction of the family of
matrices $\mathbf{C}$ is not difficult: it
suffices to impose that $\mathbf{C}$
transforms the fixed points of $\mathbf{M} $
into the ones of $\mathbf{K}_{\mathrm{C}}
(\vartheta)$, $\mathbf{A}_{\mathrm{C}} (\xi)$,
or $\mathbf{N}_{\mathrm{C}} (\nu)$, respectively.
Just to work out an example that will play a
relevant role in the forthcoming, we consider
a matrix $\mathbf{M}$ representing an elliptic
action with one fixed point denoted by $Q_f$.
Since the fixed point for the corresponding
canonical matrix $\mathbf{K}_{\mathrm{C}}
(\vartheta)$ is $+ i$, the matrix $\mathbf{C}$
we are looking for is determined by
\begin{equation}
\label{fixpoC}
\Psi[ \mathbf{C}, Q_f ] =  i .
\end{equation}
If the matrix $\mathbf{C}$ is written as
\begin{equation}
\mathbf{C} =
\left (
\begin{array}{cc}
C_1 & C_2 \\
C_3 & C_4
\end{array}
\right ) ,
\end{equation}
the solution of (\ref{fixpoC}) is
\begin{eqnarray}
\label{C2C4}
C_2 & = & - \frac{C_1 \R  Q_f +
C_3 \I Q_f}{|Q_f|^2} , \nonumber \\
& & \\
C_4 & = & \frac{C_1 \I Q_f -
C_3 \R  Q_f}{|Q_f|^2} . \nonumber
\end{eqnarray}
In addition, the condition $\det \mathbf{C}= +1$
imposes
\begin{equation}
C_3 = \sqrt{\frac{|Q_f|^2}{\I Q_f} - C_1^2} ,
\end{equation}
that, together (\ref{C2C4}) determines the matrix
$\mathbf{C}$ in terms of the free parameter $C_1$.

In figure~1.b we have plotted typical examples of
such orbits for elliptic, hyperbolic, and parabolic
actions. We stress that once the fixed points of
the ray-transfer matrix are known, one can ensure that
$Q^\prime$ will lie in the orbit associated to $Q$.

\section{First-order systems as transformations
in the Poincar\'e unit disc $\mathbb{D}$}
\label{unitdisc}

To complete the geometrical setting introduced
in the previous Section, we explore now a remarkable
transformation (introduced by Cayley) that maps
bijectively the hyperbolic plane $\mathbb{H}$
onto the unit disc, denoted by $\mathbb{D}$.
This can be done via the unitary matrix
\begin{equation}
\label{Unit}
\bm{\mathcal{U}}  =
\frac{1}{\sqrt{2}}
\left (
\begin{array}{cc}
1 & i \\
i & 1
\end{array}
\right ) ,
\end{equation}
in such a way that
\begin{equation}
\label{Cay}
\bm{\mathcal{M}} =
\bm{\mathcal{U}} \
\mathbf{M} \
\bm{\mathcal{U}}^{-1}
=
\left (
\begin{array}{cc}
\alpha  & \beta \\
\beta^\ast & \alpha^\ast
\end{array}
\right ) ,
\end{equation}
where $\bm{\mathcal{M}}$ is a matrix with
$\det \bm{\mathcal{M}} = +1$ and whose
elements are given in terms of those of
$\mathbf{M}$ by
\begin{eqnarray}
\alpha & = &  \frac{1}{2} [( A + D) +  i( C -B) ] \, ,
\nonumber \\
& & \\
\beta & = &  \frac{1}{2} [ (B + C ) + i(D - A ) ] .
\nonumber
\end{eqnarray}
In other words, the matrices $\bm{\mathcal{M}}$
belong to the group SU(1,~1), which plays an
essential role in a variety of branches in
physics. Obviously, the bilinear action induced
by these matrices is
\begin{equation}
\label{acciondis}
\mathcal{Q}^\prime  = \Phi[ \bm{\mathcal{M}},
\mathcal{Q}] = \frac{\beta^\ast +
\alpha^\ast \mathcal{Q}}
{\alpha + \beta \mathcal{Q}} ,
\end{equation}
where $\mathcal{Q}$ is the point transformed by
(\ref{Unit}) of the original $Q$:
\begin{equation}
\mathcal{Q} = \frac{Q - i}{1 - iQ} .
\end{equation}

The transformation by $\bm{\mathcal{U}}$
establishes then a one-to-one map between
the group SL(2, $\mathbb{R}$) of matrices
$\mathbf{M}$ and the group SU(1,~1) of
complex matrices $\bm{\mathcal{M}}$, which
allows for a direct translation of the
properties from one to the other.

It is easy to see that $\mathbb{H}$ maps onto
$\mathbb{D}$, as desired. The imaginary axis
in $\mathbb{H}$ goes to the $Y$ axis of the
disc $\mathbb{D}$ (in both cases, $R = \infty$
and define beam waists). In particular, $Q = +i$
is mapped onto $\mathcal{Q} = 0$. The boundary
of $\mathbb{H}$ (the real axis) goes to the
boundary of $\mathbb{D}$ (the unit circle),
and both boundaries represent fully unlimited
irradiance distributions (i.e., non-beam solutions).

Since the matrix conjugation~(\ref{Cay}) does
not change the trace, the same geometrical
classification in three basic actions still
holds. In fact, by conjugating with
$\bm{\mathcal{U}}$ the canonical forms
(\ref{Iwasa1}), we get the corresponding
ones for SU(1,~1):
\begin{eqnarray}
\label{Iwasa2}
\bm{\mathcal{K}}_{\mathcal{C}}(\vartheta )
& = & \left (
\begin{array}{cc}
\exp (i\vartheta/2) & 0 \\
0 & \exp (-i\vartheta/2)
\end{array}
\right ) \  ,
\nonumber \\
\bm{\mathcal{A}}_{\mathcal{C}} (\xi)
& = & \left (
\begin{array}{cc}
\cosh (\xi/2) & i\, \sinh(\xi/2) \\
-i\, \sinh(\xi/2) & \cosh (\xi/2)
\end{array}
\right ) \ , \\
\bm{\mathcal{N}}_{\mathcal{C}} ( \nu )
& = & \left (
\begin{array}{cc}
1 - i\, \nu/2 & \nu /2 \\
\nu /2 & 1+ i\, \nu/2
\end{array}
\right ) \  ,
\nonumber
\end{eqnarray}
that have as fixed points 0 (elliptic),
$+i$ and $-i$ (hyperbolic) and
$+i$ (parabolic), respectively. The first
matrix represent a rotation in phase space,
also called a fractional Fourier
transformation, while the second one
is sometimes called a hyperbolic
expander~\cite{SW00}.

\begin{figure}
\centering
\resizebox{0.75\columnwidth}{!}{\includegraphics{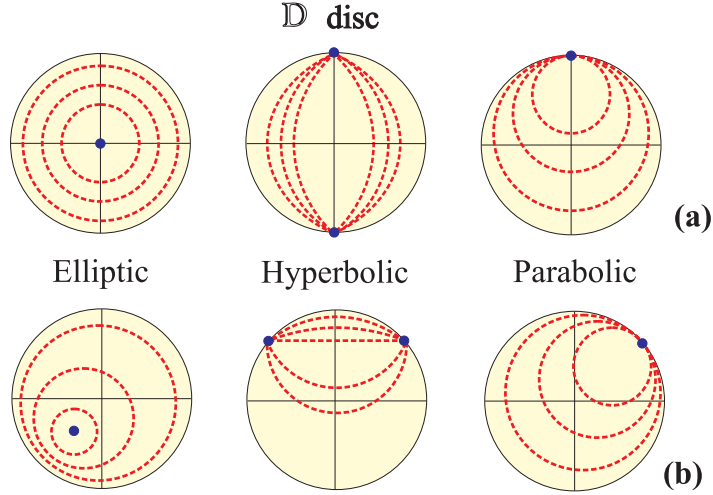}}
\caption{Plot of typical orbits in the Poincar\'e unit disc
$\mathbb{D}$: (a) canonical transfer matrices as given in
equation~(\ref{Iwasa2}) and (b) arbitrary transfer matrices.}
\end{figure}

The corresponding orbits for these matrices
are defined by
\begin{eqnarray}
\mathcal{Q}^\prime & = &
\Phi [\bm{\mathcal{K}}_{\mathcal{C}}, \mathcal{Q} ]
= \mathcal{Q}  \exp (-i\vartheta) \ , \nonumber \\
& & \nonumber \\
\mathcal{Q}^\prime & = &
\Phi [\bm{\mathcal{A}}_{\mathcal{C}}, \mathcal{Q} ]
=  \frac{\mathcal{Q} - i \tanh(\xi/2)}
{1 + i\, \mathcal{Q} \tanh(\xi/2)} \  , \\
& & \nonumber \\
\mathcal{Q}^\prime & =  &
\Phi [\bm{\mathcal{N}}_{\mathcal{C}}, \mathcal{Q} ]
= \frac{\mathcal{Q} +(1+i\mathcal{Q}) \nu/2}
{1 + (\mathcal{Q} - i) \nu /2}  . \nonumber
\end{eqnarray}

As plotted in figure~2.a, for matrices
$\bm{\mathcal{K}}_{\mathcal{C}} (\vartheta )$
the orbits are circumferences centered at the
origin. For $\bm{\mathcal{A}}_{\mathcal{C}} (\xi),$
they are arcs of circumference going from
the point $ +i$ to the point $-i$ through
$\mathcal{Q}$. Finally, for the matrices
$\bm{\mathcal{N}}_{\mathcal{C}} ( \nu )$
the orbits are circumferences passing through
the points $i$, $\mathcal{Q}$, and $-\mathcal{Q}^\ast$.
In figure~2.b we have plotted the corresponding
orbits for arbitrary fixed points.

\section{Application to optical resonators}

\label{Periodic}

The geometrical ideas presented before
allows one to describe the evolution of
a GSM beam by means of the associated
orbits.  As an application of the
formalism, we consider the illustrative
example of an optical cavity consisting
of two spherical mirrors of radii $R_1$
and $R_2$, separated a distance $d$. The
ray-transfer matrix corresponding to a
round trip can be routinely computed~\cite{GB75}
\begin{equation}
\label{CAV}
\mathbf{M} =
\left (
\begin{array}{cc}
2 g_1 g_2 - g_1 + g_2 -1 &
\displaystyle
\frac{d}{2} (2g_1 g_2 + g_1 + g_2 ) \\
\displaystyle
\frac{2}{d} (2g_1 g_2 - g_1 - g_2 ) &
2 g_1 g_2 + g_1 - g_2 - 1
\end{array}
\right ) ,
\end{equation}
where we have used the parameters ($i= 1, 2$)
\begin{equation}
g_i = 1- \frac{d}{R_i} .
\end{equation}
Note that
\begin{equation}
\label{trest}
\Tr (\mathbf{M} ) = 2 (2g_1 g_2-1) .
\end{equation}

Since the trace determines the fixed point and the
orbits of the system, the $g$ parameters establish
uniquely the geometrical action of the resonator.
To clarify further this point, in figure~3 we have
plotted the value of $|\Tr(\mathbf{M})|$ in terms
of $g_1$ and $g_2$. The plane $|\Tr(\mathbf{M})|= 2$,
which determines the boundary between elliptic
and hyperbolic action, is also shown. At the top
of the figure, a density plot is presented, with
the characteristic hyperbolic contours.

\begin{figure}
\centering
\resizebox{0.75\columnwidth}{!}{\includegraphics{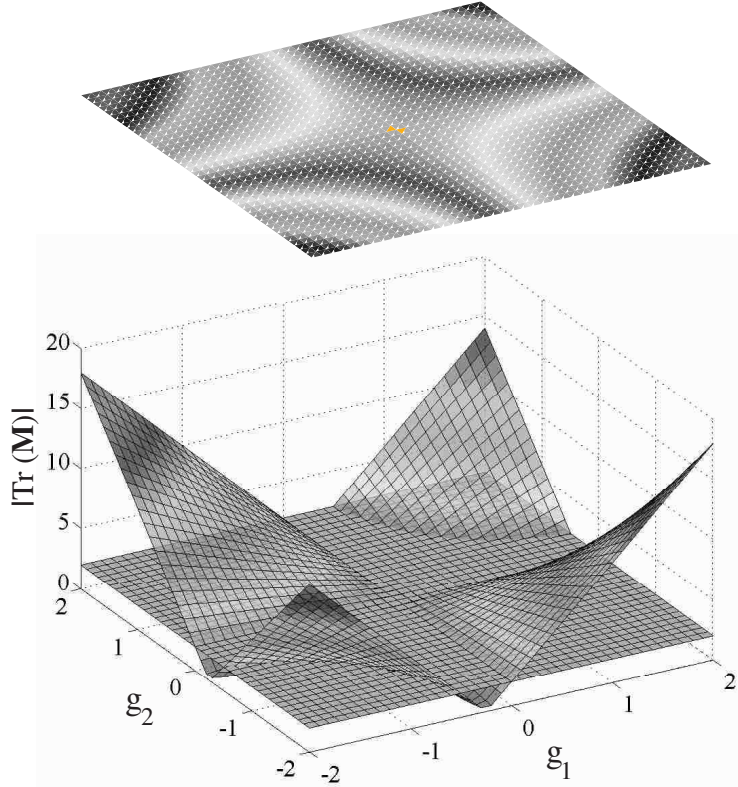}}
\caption{Plot of $|\Tr(\mathbf{M})|$ in terms of the
parameters $g_1$ and $g_2$ of the optical resonator. The plane
$|\Tr(\mathbf{M})|=2 $ is also shown. The density plot of the
three-dimensional figure appears at the bottom.}
\end{figure}

Assume now that the light bounces $N$ times
through this system. The overall transfer
matrix is then $\mathbf{M}^N$, so all the
algebraic task reduces to finding a closed
expression for the $N$th power of the matrix
$\mathbf{M}$. Although there are several
elegant ways of computing this power~\cite{Sie86},
we shall instead apply our geometrical picture:
the transformed beam is represented by the point
\begin{equation}
\label{iterat1}
Q_N = \Psi[\mathbf{M}, Q_{N-1}] =
\Psi [\mathbf{M}^N, Q_0] ,
\end{equation}
where $Q_0$ denotes the initial point.

Note that all the points $Q_N$ lie in the
orbit associated to the initial point
$Q_0$ by the single round trip, which is
determined by its fixed points: the
character of these fixed points determine
thus the behaviour of this periodic system.
By varying the parameters $g$ of the resonator
we can choose to work in the elliptic,
the hyperbolic, or the parabolic case~\cite{BK02}.

\begin{figure}[t]
\centering
\resizebox{0.75\columnwidth}{!}{\includegraphics{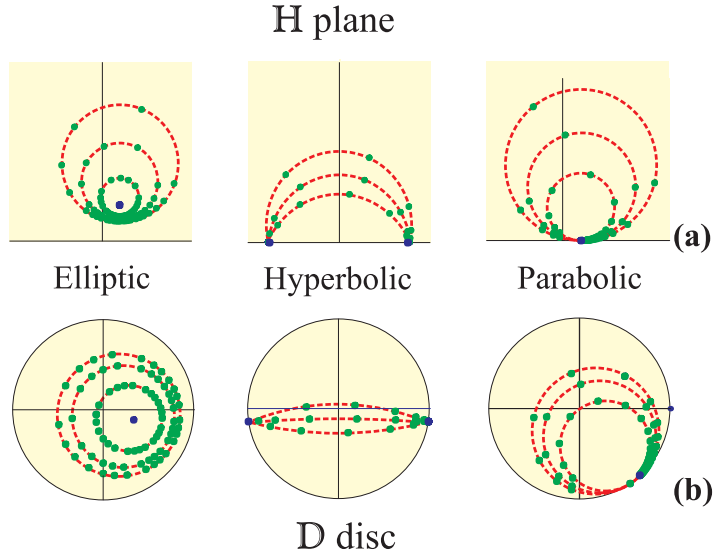}}
\caption{Plot of the successive iterates for typical elliptic,
hyperbolic, and parabolic actions. In (a) the points are
plotted in the hyperbolic plane $\mathbb{H}$, while in (b)
they are represented in the unit disc $\mathbb{D}$. For
hyperbolic and parabolic actions, the iterates tend to the
real axis and the unit circle, respectively.}
\end{figure}

To illustrate how this geometrical approach
works in practice, in figure~4.a we have plotted
the sequence of successive iterates obtained
for different kind of ray-transfer matrices,
according to our previous classification.
In figure~4.b we have plotted the same sequence
but in the unit disc, obtained via the
unitary matrix $\bm{\mathcal{U}}$.

In the elliptic case, it is clear that the
points $Q_N$ revolve in the orbit centered
at the fixed point and the system never
reaches the real axis. Equivalently, the
points $\mathcal{Q}_N$ never reach the unit
circle.

On the contrary, for the hyperbolic and
parabolic cases the iterates converge to
one of the fixed points on the real axis,
although with different laws~\cite{BMS03}.
In the general context of scattering by
periodic systems this corresponds to the
band stop and band edges,
respectively~\cite{Lek87,Yeh88,Grif01,Spru04,SCB05}.

What we conclude from this analysis is that
the iterates of hyperbolic and parabolic actions
produce solutions fully unlimited, which are
incompatible with our ideas of a beam. The only
beam solutions are thus generated by elliptic
actions and, according with equation~(\ref{trest}),
the stability criterion is
\begin{equation}
\label{stco}
0 \le | 2g_1 g_2-1 | = | \cos (\vartheta/2) | \le 1 ,
\end{equation}
where $\vartheta$ is the parameter in the canonical
form $\mathbf{K}_{\mathrm{C}}$ in equation~(\ref{Iwasa1}).
Such a condition is usually worked out in
terms of algebraic arguments using ray-transfer
matrices, although the final results apply
exclusively to scalar wave fields.

Finally, we stress that real cavities resonate
with vector fields. The situation then is far
more involved because the vector diffraction for
(polarized) electric fields is more difficult
to handle, even for systems with small Fresnel
numbers and the $ABCD$ law does not apply to the
corresponding kernel~\cite{HB94}. Exact solutions
for these vector beams have recently appeared~\cite{Lek01}.
In any case, there is abundant evidence that the stability
condition (\ref{stco}) works well. This could be
expected, since the transition to scalar theories
captures all the essential physics embodied in
the more elaborated vector analogues~\cite{MW95}.

\section{Concluding remarks}

In this paper, we have provided a geometrical
scenario to deal with first-order optical systems.
More specifically, we have reduced the action
of any system to a rotation, a translation or a
parallel displacement, according to the magnitude
of the trace of its ray-transfer matrix. These
are the basic isometries of the hyperbolic plane
$\mathbb{H}$ and also of the Poincar\'e unit
disc $\mathbb{D}$. We have also provided an
approach for a qualitative examination of the
stability condition of an optical resonator.

We hope that this approach will complement
the more standard algebraic techniques and
together they will help to obtain a better
physical and geometrical feeling for the
properties of first-order optical systems.

\end{document}